\documentclass[aps,prb,twocolumn,superscriptaddress,showpacs,floatfix]{revtex4-1}
\usepackage{graphicx,amssymb}
\usepackage{color,ulem} 

\newcommand{\cblue}[1]{{\color{blue} #1}}

\usepackage{graphicx,bm,amssymb,amsmath}
\usepackage{ulem} 

\newcommand{\bk}{{\bf k}}

\newcommand{\bq}{{\bf q}}
\newcommand{\br}{{\bf r}}

\newcommand{\bu}{{\bf u}}

\newcommand{\bw}{{\bf w}}

\def\tensor#1{\protect\@ontopof{#1}{\leftrightarrow}{1.15}\mathord{\box2}}

\begin{document}
\title{Crossover in the electron-phonon heat exchange in layered nanostructures}
\author{Dragos-Victor Anghel }
\affiliation{Horia Hulubei National Institute for R\&D in Physics and Nuclear Engineering, M\u agurele, Romania}
\author{Claudiu Caraiani}
\affiliation{University of Bucharest, Faculty of Physics, M\u agurele, Romania}
\author{Yuri M. Galperin}
\affiliation{Department of Physics, Unverstity of Oslo, PO Box 1048 Blindern, 0316 Oslo, Norway and A.~F.~Ioffe Physico-Technical Institute of Russian
Academy of Sciences, 194021 St. Petersburg, Russia }

\begin{abstract}
We study theoretically the effect of the effective dimensionality of the phonon gas distribution on the heat exchange between electrons and phonons in layered nanostructures.
If we denote the electrons temperature by $T_e$ and the phonons temperature by $T_{ph}$, then the total heat power $P$ is proportional--in general--to $T_e^x - T_{ph}^x$, the exponent $x$ being dependent on the  effective dimensionality of the phonon gas distribution.
If we vary the temperature in a wide enough range, the effective dimensionality of the phonon gas distribution changes going through a crossover around some temperature, $T_C$.
These changes are reflected by a change in $x$.
On one hand, in a temperature range well below a crossover temperature $T_C$ only the lowest branches of the phonon modes are excited.
They form a (quasi) two-dimensional gas, with $x=3.5$.
On the other hand, well above $T_C$, the phonon gas distribution is quasi three-dimensional and one would expect to recover the three dimensional results, with $x = 5$.
But this is not the case in our layered structure.
The exponent $x$ has a complicated, non-monotonous dependence on temperature forming a ``plateau region'' just after the crossover temperature range, with $x$ between 4.5 and 5.
After the plateau region, $x$ decreases, reaching values between 3.5 and 4 at the  highest temperature used  in our numerical calculations,  which is more than 40 times higher than $T_C$.
\end{abstract}
\date{October 3, 2018}

\maketitle
\section{Introduction} \label{sec_intro}

Continuous efforts in miniaturizing devices require  better understanding of their quantum behaviors at nanoscale lengths and sub-Kelvin temperatures.  Such devices are, e.g., microrefrigerators, microbolometers, and microcalorimeters.~\cite{RevModPhys.78.217.2006.Giazotto}
The microrefrigerators have been proven to be able to reduce the temperature of the electron gas of a normal metal island down to 30~mK, starting from a bath temperature of 150~mK.~\cite{PhysRevApplied.2.054001.2014.Nguyen}
Such experimental achievement allows the use of these devices in refrigerating microbolometers, microcalorimeters, qubits, etc.

The devices are usually layered structures consisting of thin metal films (for example, Cu, Al, etc.) deposited on free standing dielectric membranes, typically made of silicon nitride (SiN).
The membrane is itself supported by a bulk substrate, which may be considered the thermostat for the whole device.
A central issue in the understanding of how the devices function or how they respond to the absorption of electromagnetic radiation is the interaction and the heat exchange between the electron system and the phonon system in the normal metal films.
Our study is based on the model presented in Refs.~\onlinecite{SolidStateCommun.227.56.2016.Anghel, EurPhysJB.90.260.2017.Anghel}, in which a Cu layer of thickness $d$ is deposited on a SiN membrane of thickness $L-d$, such that the total thickness of the system is $L$.
The temperature range in which the microcalometers operate is several hundreds of mK.~\cite{PhysRevApplied.2.054001.2014.Nguyen, RevModPhys.78.217.2006.Giazotto}
For typical values of $L$, which is of the order of 100~nm, and the sound velocities of the SiN, the dominant phonon wavelength of a three-dimensional (3D) phonon gas model is comparable to $L$ at temperatures of the order of 100~mK.
As the temperature decreases, the 3D dominant phonon wavelength may become much bigger than $L$ and excitations of phonon modes perpendicular to the membrane surfaces become very unlikely.
In such a case, we can consider the phonon gas as being two-dimensional (2D).
In the other extreme, at high enough temperatures, the dominant phonon wavelength may become much smaller than $L$ and the phonon gas is practically 3D.~\cite{PhysRevLett.81.2958.1998.Anghel, PhysRevB.59.9854.1999.Anghel, PhysRevB.70.125425.2004.Kuhn, PhysRevB76.165425.2007.Kuhn}


At temperatures of the  order of 1~K, due to the depopulation of the phonon modes and the degeneracy of the electrons system, the two subsystems of a normal metal--the electrons and the phonons--become almost isolated from each other(see, e.g., the discussion in Sec.~II of Ref.~\onlinecite{RevModPhys.78.217.2006.Giazotto}).
For this reason, they can ``equillibrate'' independently, that is, they may reach equilibrium distributions over the quasiparticle states at some effective temperatures that we shall denote by $T_e$ (for electrons) and $T_{ph}$ (for phonons).\footnote{
Note that we don't aim at \textit{calculation} of the temperatures $T_e$ and $T_{ph}$, which may differ from the temperature of bulk substrate (the thermostat).
Therefore we do not need to include phonon-electron scattering explicitly.
}
Using the effective temperatures, the heat exchange $P$ between the electrons and the phonons may be written in most cases as \cite{PhysRevB.49.5942.1994.Wellstood, PhysRevB.81.245404.2010.Viljas, PhysRevB.93.115405.2016.Cojocaru, PhysRevB.77.033401.2008.Hekking}
%
\begin{subequations} \label{Ps_ansatz}
\begin{equation}
  P = V \Sigma (T_e^x - T_{ph}^x), \label{P_ansatz}
\end{equation}
where $V$ is the volume and $\Sigma$ is a constant of material.
The exponent $x$ varies, depending on the model and the dimensionalities of the electrons and the phonon gases.
For 
3D models, $x = 5$,~\cite{PhysRevB.49.5942.1994.Wellstood}  for 
quasi-2D ones, $x = 4$,\cblue{\cite{PhysRevB.81.245404.2010.Viljas}} whereas for a 1D phonon gas  coupled to a 3D electrons system, $x = 3$.\cite{PhysRevB.77.033401.2008.Hekking}
Apparently, this leads to the conjecture
\begin{equation}
  x = 2 + s  \label{s_ansatz}
\end{equation}
\end{subequations}
where $s$ is the dimensionality of the phonon gas.
Apart from Eq.~(\ref{s_ansatz}), it was observed that $s$ may take also fractional values: if the phonon gas is 2D and the electron gas distribution is a superposition of 2D conduction bands, then $x=3.5$.~\cite{SolidStateCommun.227.56.2016.Anghel, EurPhysJB.90.260.2017.Anghel}

The exponent $x$ depends on the effective dimensionality of the phonon gas, but the latter may change with temperature, as we discussed above.~\cite{PhysRevLett.81.2958.1998.Anghel, PhysRevB.59.9854.1999.Anghel, PhysRevB.70.125425.2004.Kuhn, PhysRevB76.165425.2007.Kuhn}
Therefore, in a temperature range where the dimensionality of the phonon gas changes, the exponent $x$ may also change, showing a crossover between the values corresponding to different dimensionalities.
This issue, which, to the best of our knowledge, has not been investigated will be explored here, as a continuation of previous  studies.~\cite{SolidStateCommun.227.56.2016.Anghel, EurPhysJB.90.260.2017.Anghel}

The paper is organized as follows.
In the next section we present the theory, with the system's specifications, then, we present the results, and finally, we draw the conclusions.
In the following, we try to preserve the notations from Refs.~\onlinecite{SolidStateCommun.227.56.2016.Anghel,EurPhysJB.90.260.2017.Anghel}
as much as possible.

\section{Theory
}
\label{sec_method}

As in Refs.~\onlinecite{SolidStateCommun.227.56.2016.Anghel, EurPhysJB.90.260.2017.Anghel}, we consider a metallic film (for example, Cu) deposited on a free-standing dielectric membrane (for example silicon nitride) and perfectly glued to it.~\cite{ApplPhysLett.82.293.2003.Anghel, ApplPhysLett.78.556.2001.Anghel}
For the mathematical description, we choose a Cartesian system of coordinates, such that all the surfaces of the membrane and of the film are parallel to the $(xy)$ plane.
In the $z$ direction, the membrane extends in the interval $[-L/2, L/2-d]$, whereas the metal occupies the interval $[L/2-d, L/2]$.
In the $x$ and $y$ directions, the system is considered to be very large (in comparison with the wavelengths of the quasiparticles involved) and has the area $A$;
we assume that
$L \ll \sqrt{A}$.

\subsection{Description of the electrons}

The metal contains free electrons, which interact with the phonons that propagate in the whole system.
The electrons' wave functions propagate in the $(xy)$ plane and are 
confined
along the $z$ direction.
We denote the electron's wavevector by $\bk \equiv (\bk_\parallel, k_z)$, where $\bk_\parallel$ and $k_z$ are the components of $\bk$ perpendicular and parallel to the $z$ axis, respectively.
For $\bk_\parallel$, we impose periodic boundary conditions in the $(xy)$ plane, whereas for $k_z$ we impose Dirichlet boundary conditions at $z = L/2 - d$ and $z = L/2$.
This leads to a density of (allowed quantum) states (DOS) in the $\bk_\parallel$ variable equal to $A/(2\pi)^2$ and to the quantization condition $k_z \equiv n\pi/d$, where $n$ is an integer.
Then,  the electron's wavefunction is of the form 
$$\psi _{\bk_\parallel,n}(\br,t) \equiv \psi _{\bk_\parallel,k_z}(\br,t) = \phi _{k_{z}}(z)e^{i(\bk_\parallel \br_{\parallel }-\epsilon_{\bk_\parallel,n}t/\hbar )}/\sqrt{A}, $$ 
where
$\phi _{k_{z}}(z)=\sqrt{\frac{2}{d}}\sin \left[ \left( z+d-\frac{L}{2}\right)
k_{z}\right]$ 
and
the electron's energy is $$\epsilon _{\mathbf{k}}=\frac{\hbar ^{2}k^{2}}{2m_{e}}=\frac{\hbar^{2}k_{\parallel }^{2}}{2m_{e}}+\frac{\hbar ^{2}k_{z}^{2}}{2m_{e}}\equiv \epsilon _{k_{\parallel },k_{z}}, $$ where $m_{e}$ is the free electron mass.
In these notations, the electron annihilation and creation field operators are
\begin{subequations}
\begin{eqnarray}
  \Psi (\mathbf{r},t) & = & \sum_{\bk_\parallel, k_{z}} \psi _{\bk_\parallel, k_{z}}(\br, t) c_{\bk_\parallel, k_{z}}\\
   \Psi ^{\dag }(\mathbf{r},t) & = & \sum_{\mathbf{k}_{\parallel },k_{z}}\psi _{\bk_\parallel, k_{z}}^* (\br,t) c_{\mathbf{k}_{\parallel},k_{z}}^\dag,  \label{def_Psi}
\end{eqnarray}
\end{subequations}
respectively, where $c_{\mathbf{k}_{\parallel },k_{z}}$ and $c_{\bk_\parallel, k_{z}}^{\dag}$ are the electron annihilation and creation operators on the state $\psi _{\bk_\parallel,k_z}$.

Since $\sqrt{A} \gg L > d$, $\bk_\parallel \equiv (k_x, k_y)$ has a quasi-continuous spectrum, whereas $k_z$ will be considered discreet.
States of constant $k_z$ (or $n$) and any value of $k_\parallel$ form the 2D \textit{conduction bands}.
We denote by $n_F$ the number of occupied 2D bands at $T_e = 0$~K,
\begin{equation}
  n_F \equiv \left\lfloor \frac{\sqrt{2m_eE_F}}{\pi \hbar} d \right\rfloor ,
  \label{def_nF}
\end{equation}
where $\lfloor x\rfloor$ is the biggest integer smaller or equal to $x$ and $E_F$ is the Fermi energy.
If the normal metal is Cu, $E_F \approx 7$~eV and for $d = 10$~nm,  $n_F=43$, whereas for $d=20$~nm, $n_F= 86$.

\subsection{Description of the phonons}

The phonons are quanta of elastic vibrations of the system.
They propagate in the $(xy)$ plane and are stationary waves along $z$.
At the surfaces of the system, at $z = \pm L/2$, the phonons satisfy free boundary conditions.
To simplify the calculation of the phonon modes,~\cite{SolidStateCommun.227.56.2016.Anghel, EurPhysJB.90.260.2017.Anghel}  we consider that the metal and the dielectric membrane have identical mechanical properties, namely they have the same mass density and identical elastic properties.
Although this seems to be a crude simplification, it gives us a method to study the characteristics of the electron-phonon interaction in general. 

The phonon modes are eigenstates of the elastic dynamic equations and are grouped in three polarizations: horizontal shear ($h$), symmetric ($s$) and antisymmetric ($a$) modes~\cite{Auld:book}.
While the $h$ modes are pure transversal ($t$) vibrations, the $s$ and $a$ modes are superpositions of transversal and longitudinal ($l$) vibrations~\cite{Auld:book}--in a longitudinal vibration, the displacement field is in the propagation direction, whereas in a transversal vibration the displacement field and the propagation direction are perpendicular to each other.

An $h$ mode is characterized by a single wavevector, of components $(\bq_\parallel, q_t)$, perpendicular and parallel to the $z$ axis, respectively.
The component $\bq_\parallel$ satisfy periodic boundary conditions in the $(xy)$ plane and $q_t \equiv \nu\pi/L$, where $\nu = 0, 1, \ldots$.
States of constant $\nu$ and variable $\bq_\parallel$ define 2D \textit{phonon branches}.

An $s$ mode--and, similarly, an $a$ mode--is characterized by two wavevectors, corresponding to the $l$ and $t$ vibrations.
The wavevector corresponding to the longitudinal vibration has the components $\bq_\parallel$ and $q_l$, whereas the wavevector of the transversal vibration has the components $\bq_\parallel$ and $q_t$.
The component $\bq_\parallel$, perpendicular to the $z$ axis, is the same for both, $l$ and $t$ vibrations, whereas the components $q_l$ and $q_t$, parallel to the $z$ axis, are, in general, different.
As in the case of electrons, $\bq_\parallel$ satisfies periodic boundary conditions on the area $A$ and has a DOS of $A/(2\pi)^2$.
The free boundary conditions imposed at $z = \pm L/2$ lead to the equation~\cite{Auld:book, EurPhysJB.90.260.2017.Anghel}
\begin{subequations} \label{syst_qtql}
\begin{equation}
  \frac{- 4q_\parallel^2 q_{l\xi}q_{t\xi}}{(q_\parallel^2-q_{t \xi}^2)^2} = \left[ \frac{\tan(q_{t\xi}L/2)} {\tan(q_{l\xi}L/2)} \right]^{\pm 1} ,  \label{6}
\end{equation}
for the components $q_l$ and $q_t$, where $q_\parallel \equiv |\bq_\parallel|$.
The exponents $+1$ and $-1$ on the right hand side (rhs) of Eq.~(\ref{6}) correspond to the polarizations $s$ and $a$, respectively~\cite{Auld:book}, whereas $\xi \equiv (\alpha, \nu)$ is a doublet, containing the polarization $\alpha = s, a$ and the branch number $\nu = 0,1,2, \ldots$ (as in the case of the $h$ polarization), as we shall specify below.
If we denote by $c_l$ and $c_t$ the longitudinal sound velocity and the transversal sound velocity, respectively, then the angular frequency, common to both, longitudinal and transversal vibrations, is given by
%
\begin{equation}
\omega_{q_\parallel\xi}=c_{l}\sqrt{q_{l,\xi}^{2}+q_\parallel^{2}}=c_{t} \sqrt{q_{t,\xi}^{2}+q_\parallel^{2}} .  \label{8}
\end{equation}
\end{subequations}
The system~(\ref{syst_qtql}) has a countable, infinite number of solutions, for each $\alpha$ and $q_\parallel$.
These solutions are denoted by $\nu$ and form the branches, when plotted as functions of $q_\parallel$.
In general, $q_{t\alpha\nu}$ and $q_{l\alpha\nu}$ may take both, (positive) real and imaginary values, as explained for example in \cite{Auld:book} (see also \cite{JPhysA.40.10429.2007.Anghel}), whereas $q_\parallel$ takes only real positive values, by definition.
If $\rho$ is the mass density of the system (considered homogeneous in the whole volume), then the sound velocities may be expressed in terms of the Lam\'e coefficients $\lambda$ and $\mu$~\cite{Auld:book},
\begin{equation}
  c_{t}^{2}=\frac{\mu }{\rho },\quad c_{l}^{2}=\frac{\lambda +2\mu }{\rho } , \label{ct_cl}
\end{equation}
and we define $J \equiv c_{t}^{2}/c_{l}^{2}$.

We define the elastic modes by 
$\bw_{\bq_\parallel \xi}(z) e^{i (\bq_\parallel \br_\parallel - \omega_{\bq_\parallel \xi} t)} /\sqrt{A}$,  where the functions $\bw_{\bq_\parallel \xi}(z)$
are normalized such that $\int_{-L/2}^{L/2} \bw_{\bq_\parallel \xi}(z)^{\dag} \bw_{\bq_\parallel \xi'}(z)dz = \delta_{\xi,\xi'}$ and have the following analytical expressions,~\cite{SolidStateCommun.227.56.2016.Anghel, EurPhysJB.90.260.2017.Anghel}
\begin{subequations} \label{def_wsa_xz}
\begin{eqnarray}
  w_{\bq_\parallel,s,\nu,x} &=& N_s iq_{t}\left[  2q_\parallel^{2}\cos\left(  \frac{q_{t}L}{2}\right)  \cos\left(q_{l}z\right)  +(q_{t}^{2}-q_\parallel^{2}) \right. \nonumber \\
  && \left. \times \cos\left(  \frac{q_{l}L}{2}\right) \cos\left(  zq_{t}\right)  \right]  , \label{def_ws_x} \\
  w_{\bq_\parallel,s,\nu,z} &=& N_s q_\parallel\left[  -2q_{t}q_{l}\cos\left(  \frac{q_{t}L}{2}\right) \sin\left(  q_{l}z\right)  +(q_{t}^{2}-q_\parallel^{2}) \right. \nonumber \\
  && \left. \times \cos\left(\frac{q_{l}L}{2}\right)  \sin \left( zq_{t}\right)  \right]  , \label{def_ws_z} \\
  w_{\bq_\parallel,a,\nu,x} &=& N_a iq_{t}\left[  2q_\parallel^{2}\sin\left(  \frac{q_{t}L}{2}\right)  \sin\left(q_{l}z\right) + (q_{t}^{2}-q_\parallel^{2}) \right. \nonumber \\ 
  && \left. \times \sin\left(  \frac{q_{l}L}{2}\right) \sin\left(  zq_{t}\right)  \right]  , \label{def_wa_x} \\
  w_{\bq_\parallel,a,\nu,z} &=& N_a q_\parallel\left[  2q_{t}q_{l}\sin\left(  \frac{q_{t}L}{2}\right)  \cos\left(q_{l}z\right) - (q_{t}^{2}-q_\parallel^{2}) \right. \nonumber \\
  && \left. \times \sin \left(\frac{q_{l}L}{2}\right) \cos\left(  zq_{t}\right)  \right]  , \label{def_wa_z}
\end{eqnarray}
\end{subequations}
with the normalization constants
\begin{widetext}
\begin{subequations} \label{defs_NsNa}
\begin{eqnarray}
\frac{1}{N_s^{2}} & = & A\bigg\{4q_t^2q_\parallel^2\cos^2(q_tL/2)\bigg[(p_l^2+q_\parallel^2)\frac{\sinh(p_lL)}{2p_l} -(p_l^2-q_\parallel^2)\frac{L}{2}\bigg] 
\nonumber \\ & &
+(q_t^2-q_\parallel^2)^2\cosh^2(p_lL/2)\bigg[(q_t^2+q_\parallel^2)\frac{L}{2} + (q_t^2 - q_\parallel^2)\frac{\sin(q_tL)}{2q_t}\bigg] 
-4q_\parallel^2q_t(q_t^2-q_\parallel^2)\cosh^2\left(\frac{p_lL}{2}\right)\sin(q_tL)\bigg \} , \label{def_Ns} \\
\frac{1}{N_a^{2}} & = & A\bigg\{4p_t^2q_\parallel^2\sinh^2(p_tL/2)\bigg[(p_l^2+q_\parallel^2)\frac{\sin(p_lL)}{2p_l} +(p_l^2-q_\parallel^2)\frac{L}{2}\bigg] 
\nonumber \\ & &
+(p_t^2-q_\parallel^2)^2\sinh^2(p_lL/2)\bigg[(p_t^2+q_\parallel^2)\frac{\sinh(p_tL)}{2} - (p_t^2 - q_\parallel^2)\frac{L}{2p_t}\bigg]
-4q_\parallel^2p_t(p_t^2+q_\parallel^2)\sinh^2\! \left(\!\frac{p_lL}{2}\!\right)\sinh(p_tL)
\!  \bigg \} \! . \label{def_Na}
\end{eqnarray}
\end{subequations}
\end{widetext}
Then the displacement field operator is
\begin{eqnarray}
  \bu (\br, t) &=& \sum_{\xi,\bq_\parallel} \sqrt{\frac{\hbar}{2 \rho \omega_{\bq_\parallel \xi} }}
   e^{i (\bq_\parallel \br_\parallel - i\omega_{\bq_\parallel \xi}t )}
   \nonumber \\ && \times
   \left[ a_{\bq_\parallel \xi} \bw_{\bq_\parallel \xi}(z) + a_{-\bq_\parallel \xi}^{\dag}  \bw^*_{\bq_\parallel \xi}(z) \right] ,
\label{displ_operator}
\end{eqnarray}
where $a_{\bq_\parallel \xi}^\dag$ and $a_{\bq_\parallel \xi}$ are the phonon creation and annihilation operators.

In the following we shall write $q_t$ instead of $q_{t,\xi}$ and $q_l$ instead of $q_{l,\xi}$, when this does not lead to confusions.

The electron-phonon interaction is calculated in the deformation potential model, employing the interaction Hamiltonian,~\cite{Ziman:book}
\begin{equation}
  H_{\rm def}=\frac{2}{3}E_{F}\int_{V_{el}=A\times d}d^{3}\mathbf{r}\,\Psi^{\dagger}(\mathbf{r})\Psi (\mathbf{r})\nabla \cdot \mathbf{u}(\mathbf{r}) . \label{def_def_pot}
\end{equation}
Starting from Eq.~(\ref{def_def_pot}) and applying the Fermi golden rule $\Gamma_{i\to f} (2\pi/\hbar) |\langle f| H_{\rm def}| i\rangle |^2 \delta(E_f - E_i)$ to calculate the transition rates $\Gamma$ between the initial state $|i\rangle$, of energy $E_i$, and the final state $|f\rangle$, of energy $E_f$, the heat power was calculated in Refs.~\onlinecite{SolidStateCommun.227.56.2016.Anghel, EurPhysJB.90.260.2017.Anghel} to be
\begin{subequations} \label{defs_P0P1}
\begin{equation}
  P \equiv P^{(0)} \left( T_{e}\right) - P^{(1)} ( T_e, T_{ph}) ,  \label{17}
\end{equation}
where
\begin{eqnarray}
&&  P^{(0)} ( T_{e} ) \equiv P^{(0)}_s ( T_{e} ) + P^{(0)}_a ( T_{e} ) , \label{def_P0} \\
&&  P^{(1)} ( T_{e}, T_{ph} ) \equiv  P^{(1)}_s ( T_{e}, T_{ph} ) + P^{(1)}_a ( T_{e}, T_{ph} ) , \label{def_P1} \\
 &&  P^{(0)}_\alpha ( T_{e} ) \equiv  \frac{4\pi}{\hbar} \sum_{\bk_\parallel \bk_\parallel', n, n'}^{\bq_\parallel, \nu} \hbar \omega _{\bq_\parallel, \alpha, \nu} |g_{\alpha, \nu,\bq_\parallel}^{n',n}|^{2}
  \nonumber \\ && \quad \times
   [f(\beta_e \epsilon_{\mathbf{k}_\parallel -\mathbf{q}_\parallel, n'}) - f(\beta_e \epsilon_{k_\parallel,n}) ]  
  n(\beta_e \epsilon_{q_\parallel, \nu}) , \label{def_P0a} \\
&&  P^{(1)}_\alpha ( T_e, T_{ph}) \equiv \frac{4\pi}{\hbar} \sum_{\bk_\parallel \bk_\parallel', n, n'}^{\bq_\parallel, \nu} \hbar \omega _{\bq_\parallel, \alpha, \nu} |g_{\alpha, \nu,\bq_\parallel}^{n',n}|^{2} 
\notag \\ && \quad \times
[f(\beta_e \epsilon_{\mathbf{k}_\parallel -\mathbf{q}_\parallel, n'}) - f(\beta_e \epsilon_{k_\parallel,n}) ]  
  n(\beta_{ph} \epsilon_{q_\parallel, \nu}) , \label{def_P1a} 
\end{eqnarray}
\end{subequations}
whereas $\alpha = s$ or $a$.~\cite{SolidStateCommun.227.56.2016.Anghel, EurPhysJB.90.260.2017.Anghel}
We also used the notation
\begin{eqnarray}
  g_{\xi ,\bq_{\parallel }}^{n',n} &=& \frac{2}{3} E_{F} N_{\xi, q_\parallel} \sqrt{\frac{\hbar }{2\rho \omega _{\bq_\parallel, \xi}}} \int_{L/2-d}^{L/2}\phi _{n^{\prime }}^* (z)\phi _{n}(z)  \notag \\
  && \times \left[ i\bq_\parallel \cdot \bw_{\bq_\parallel, \xi}(z)+\frac{d w_{\bq_\parallel, \xi}(z)}{d z}\right] dz,  \label{3a}
\end{eqnarray}
where the normalization constants $N_{\xi, q_\parallel}$ are given in Eqs.~(\ref{defs_NsNa}).

With the power defined by Eqs.~(\ref{defs_P0P1}), the exponents of the temperature dependence, as defined in~(\ref{P_ansatz}), are calculated as
\begin{subequations}\label{defs_xTe_xTph}
\begin{equation}
  x_{T_{ph}} \equiv - \frac{\partial \ln(P)}{\partial \ln(T_{ph})} \quad {\rm and} \quad x_{T_e} \equiv \frac{\partial \ln(P)}{\partial \ln(T_e)} . \label{defs_xTe_xTph1}
\end{equation}
Although in Eq.~(\ref{P_ansatz}) we have only one exponent $x$, in Eq.~(\ref{defs_xTe_xTph}) we introduce both, $x_{T_{ph}}$ and $x_{T_e}$, to check the consistency of the description.

If $P^{(1)}$ does not depend on $T_e$ (as we shall see further) and since $P^{(0)}$ does not depend on $T_{ph}$, then we may write Eqs.~(\ref{defs_xTe_xTph1}) as
\begin{equation}
  x_{T_{ph}} = - \frac{\partial \ln(P^{(1)})}{\partial \ln (T_{ph})} \quad {\rm and} \quad x_{T_e} = \frac{\partial \ln(P^{(0)})}{\partial \ln(T_e)} . \label{defs_xTe_xTph2}
\end{equation}
\end{subequations}

\section{Results} \label{sec_results}

For concreteness, we consider systems of total thickness $L = 100$~nm, of which, the thickness of the metallic film is around 10~nm.
The elastic properties of the entire system correspond to the silicon nitride, which is a common material for the construction of the supporting membrane of nano-detectors.~\cite{RevModPhys.78.217.2006.Giazotto}
For this material, the density is $\rho = 3290~{\rm kg/m^3}$, whereas the longitudinal and transversal sound velocities are $c_l = 10300$~m/s and $c_t = 6200$~m/s, respectively.
The crossover temperature is defined as $T_C = c_t \hbar/(2 k_B L) \approx 237$~mK,~\cite{PhysRevB.70.125425.2004.Kuhn} and in the low temperature limit ($T \ll T_C$) only the lowest $s$ and $a$ phonon branches contribute to the electron-phonon interaction, since the upper branches are depopulated.~\cite{SolidStateCommun.227.56.2016.Anghel, EurPhysJB.90.260.2017.Anghel}
As the temperature is increasing, the phonon gas changes dimensionality in a temperature interval around $T_C$ and upper $s$ and $a$ branches gradually start to be populated and play a role in the interaction.

In the electron system, all the bands, from 1 to $n_F$, are populated at any temperature and contribute -- with the electrons close to the Fermi energy -- to the electron-phonon interaction.
To estimate the difference between the energies of electrons of the same $k_\parallel$, but belonging to different 2D conduction bands, for a metallic layer of thickness $d = 10$~nm, we calculate $(\epsilon_{\mathbf{k}_\parallel = \mathbf{0}, n=2} - \epsilon_{\mathbf{k}_\parallel = \mathbf{0},n=1})/k_B \approx 131$~K and $(\epsilon_{\mathbf{k}_\parallel = \mathbf{0}, n=44} - \epsilon_{\mathbf{k}_\parallel = \mathbf{0}, n=43})/k_B \approx 3796$~K.
We observe that this difference is very large in comparison with the energy of the thermal phonons, so scattering of electrons between different conduction bands are very rare and therefore they do not contribute to the electron-phonon heat exchange in a temperature range below 10~K (see also~\cite{SolidStateCommun.227.56.2016.Anghel, EurPhysJB.90.260.2017.Anghel}).

\begin{figure}[t]
  \centering
  \includegraphics[width=7cm,keepaspectratio=true]{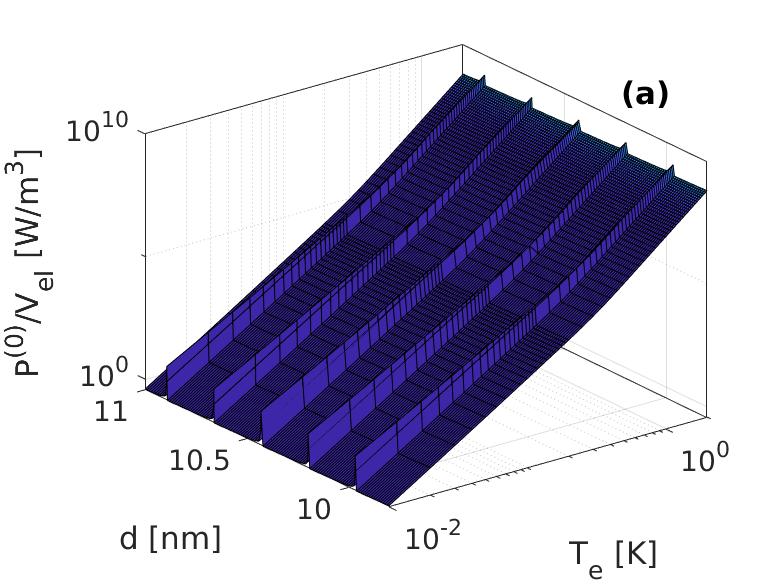}
  \includegraphics[width=7cm,keepaspectratio=true]{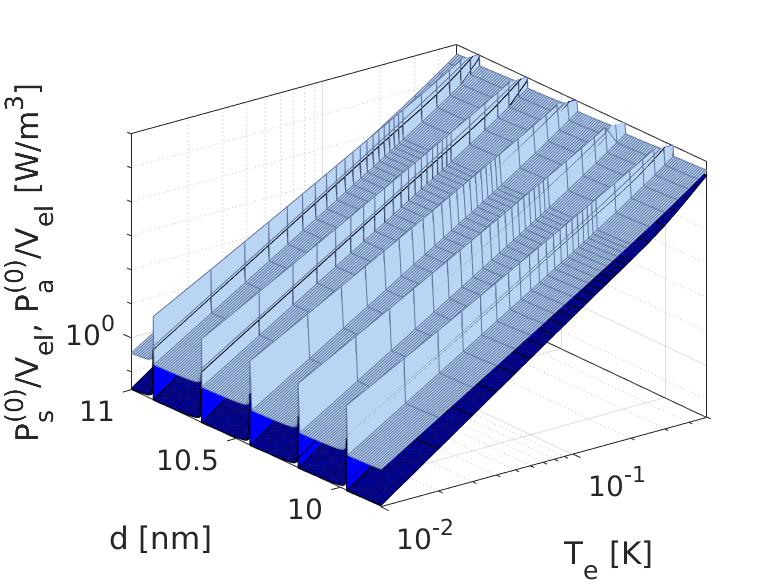}
  \caption{(a) The heat power $P^{(0)}$, as a function of $T_e$ and the thickness $d$ of the metallic layer. In (b) we show separately (and on a smaller $T_e$ range, for clarity) the contribution of the symmetric modes $P^{(0)}_s$ (dark blue, below) and of the antisymmetric modes $P^{(0)}_a$ (light blue, above). In the low temperature limit, $P^{(0)}_s < P^{(0)}_a$ for any $d$.}
  \label{P0s_P0a_Ptot}
\end{figure}

\begin{figure}[t]
  \centering
  \includegraphics[width=7cm,keepaspectratio=true]{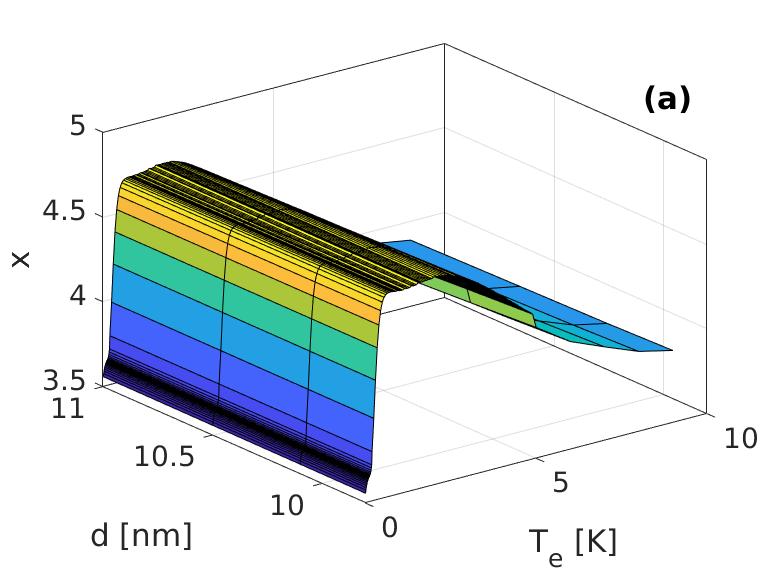}
  \includegraphics[width=7cm,keepaspectratio=true]{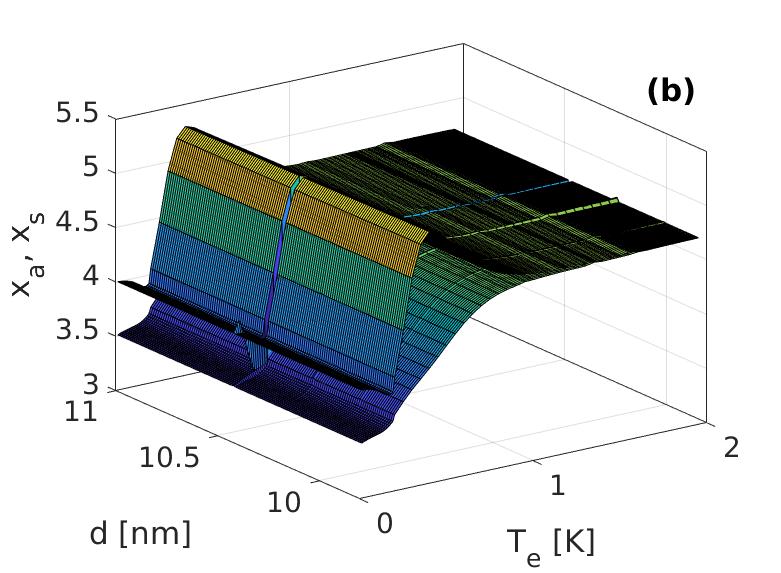}
  \caption{The exponent of the temperature dependence.}
  \label{exes}
\end{figure}

In Fig.~\ref{P0s_P0a_Ptot}~(a) we plot $P^{(0)}$ vs $T_e$ and $d$.
We observe the formation of the narrow crests, separated by valleys, as in Refs.~\onlinecite{SolidStateCommun.227.56.2016.Anghel, EurPhysJB.90.260.2017.Anghel}.
Both, $T_e$ and $P^{(0)}/V_{el}$ are plotted in logarithmic scale and we observe a change of the slope of $\ln P^{(0)}$ vs $\ln T_e$, indicating a dimensionality crossover of the  phonon gas distribution.
Apparently, the slope is increasing with temperature, but more details are seen in Fig.~\ref{exes}, where we show a larger temperature interval.
In Fig.~\ref{P0s_P0a_Ptot}~(b) we plot separately the contribution of the symmetric phonon modes $P^{(0)}_s$ (lower values in the low temperature limit) and of the antisymmetric phonon modes $P^{(0)}_a$ (higher values in the low temperature limit).

Similarly, we calculated $P^{(1)}_s (T_e, T_{ph})$, $P^{(1)}_a (T_e, T_{ph})$, and  $P^{(1)} (T_e, T_{ph}) \equiv P^{(1)}_s (T_e, T_{ph}) + P^{(1)}_s (T_e, T_{ph})$.
In the valleys (between crests) and in the temperature range investigated by us, both, $P^{(1)}_s$ and $P^{(1)}_a$ are independent of $T_e$, whereas their dependence on $T_{ph}$ is the same with that of $P^{(0)}_s$ and $P^{(0)}_a$ on $T_e$, within the numerical accuracy.
Introducing a notation, $T_x$, for both, $T_e$ and $T_{ph}$, and a constant value $T_{e0}$, we may write, formally,
\begin{equation}
  P^{(1)}_s (T_{e0}, T_x) = P^{(0)}_s(T_x) \quad {\rm and} \quad
  P^{(1)}_a (T_{e0}, T_x) = P^{(0)}_a(T_x) . \label{num_id}
\end{equation}
For this reason we do not plot also $P^{(1)}_s$, $P^{(1)}_a$ or $P^{(1)} (T_e, T_{ph})$ vs $T_{ph}$ and $d$.
Numerically, these plot would be identical with the plots in Fig.~\ref{P0s_P0a_Ptot}.  

In Fig.~\ref{exes}(a) we plot the exponent of the temperature dependence of $P^{(0)}$, denoted simply as $x$--since from Eqs.~(\ref{defs_xTe_xTph}) and (\ref{num_id}) we now know that $x_{T_e} = x_{T_{ph}}$.
At low $T_e$, $x = 3.5$ (see~\cite{SolidStateCommun.227.56.2016.Anghel, EurPhysJB.90.260.2017.Anghel}), then increases with temperature to a value between 4.5 and 5 (not to 5, as expected from~\cite{PhysRevB.49.5942.1994.Wellstood}) and varies little in an interval $T_e \in [0.5~{\rm K}, 3.5~{\rm K}]$, forming the ``plateau region.''
Quite surprisingly, at temperatures above above the plateau ($T_e > 3.5$~K) the exponent decreases again, reaching a value between 3.5 and 4, at $T_e = 10$~K.
This is surprising because $T_e=10$~K is more than 1 order of magnitude above the crossover temperature $T_C$ and we would expect that at such temperatures the value of $x$ should correspond to a 3D phonon gas $x = 5$,~\cite{PhysRevB.49.5942.1994.Wellstood} see Eq.~(\ref{s_ansatz}) and the discussion around it.

In Fig.~\ref{exes}(b) we show the exponents of the temperature dependence of both, $P^{(0)}_s$ and $P^{(0)}_a$,
\begin{eqnarray}
  x_s &\equiv& \frac{\partial \ln(P^{(0)}_s)}{\partial \ln(T_e)} = \frac{\partial \ln(P^{(1)}_s)}{\partial \ln(T_{ph})} , \notag \\ 
  x_a &\equiv & \frac{\partial \ln(P^{(0)}_a)}{\partial \ln(T_e)} = \frac{\partial \ln(P^{(1)}_a)}{\partial \ln(T_{ph})}  \label{defs_xs_xa}
\end{eqnarray}
in a temperature range between 0 and 2~K.
This is a higher resolution image, where we can notice also the crests.
Nevertheless, in the crests regions the numerical errors are big, due to the high gradients and we shall not take them into account in our discussion.
In the low temperature limit, $x_s = 4$ and $x_a = 3.5$.
We notice that $x_a$ increases from the low temperature value to a value between 4.5 and 5 in the plateau region (see above), after passing a crossover region around $T_C$.

The variation of $x_s$ is more complicated, as it can be seen in Fig.~\ref{exes}(b).
In the low temperature limit, $x_s = 4$ and then oscillates in the crossover region, until it reaches the plateau, where it becomes practically equal to $x_a$.
Notice that in the crossover region $x_s$ increases above 5 (which is supposed to be the 3D value).
At the plateau and after it, $x_s$, $x_a$, and $x$ become gradually (as temperature increases) numerically indistinguishable. For this reason, there was no point in plotting separately $x_s$ and $x_a$ at temperatures above 2~K.

\section{Conclusions} \label{sec_concl}

We studied the effect of the dimensionality of the phonon gas distribution on the heat exchange between electrons and phonons in a layered structure, consisting of a metallic layer (in our case, Cu) of thickness around 10~nm deposited on an insulating free standing membrane.
The total thickness of the system (metal plus supporting membrane) is $L = 100$~nm. 
From a mechanical point of view (elastic properties and mass density) the system  is considered homogeneous--the elastic waves propagate through a homogeneous material, making no difference between the metal and the dielectric membrane.
The longitudinal sound velocity, the transversal sound velocity, and the density of the system are considered to be those of the silicon nitride.
With these parameters, the crossover temperature around which the phonon gas distribution changes from two-dimensional (at lower temperatures) to three dimensional (at higher temperatures) is $T_C \approx 237$~mK.~\cite{PhysRevB.70.125425.2004.Kuhn}

We denoted the electron temperature by $T_e$ and the phonon temperature by $T_{ph}$
The heat  power $P$ is expected to be described by the ansatz~(\ref{Ps_ansatz}).
It was already noticed that the ansatz does not apply to our type of layered system and in the low temperature limit, the exponent corresponding to the 2D phonon gas distribution is $x = 3.5$, not 4.~\cite{SolidStateCommun.227.56.2016.Anghel, EurPhysJB.90.260.2017.Anghel}
We recovered here the results of Refs.~\onlinecite{SolidStateCommun.227.56.2016.Anghel, EurPhysJB.90.260.2017.Anghel} and we expected that, as the temperature increases and the phonon gas distribution changes from 2D to 3D, $x$ should change continuously from 3.5 to the value corresponding to 3D systems, which is 5.~\cite{PhysRevB.49.5942.1994.Wellstood}
Instead of this, we observed that $x$ does not have a monotonic behavior and totally disobeys the ansatz.
It starts from $x=3.5$ at low temperatures, it increases through the crossover region of temperatures around $T_C$ and reaches a ``plateau region'' in which $x$ varies in a small interval which lies between 4.5 and 5.
The plateau region lies in a temperature range roughly between 0.5 and 5~K.
After this, the exponent decreases steadily reaching values between 3.5 and 4 at the upper limit of the temperature range investigated by us, which is 10~K.

The results did not confirm the behavior of the heat power in the corresponding limits, as expected from the literature.
Nevertheless, the exponent $x$ should convey valuable information regarding the dispersion relations and the dimensionalities of the phonon system and of the electrons system.
More analytical work is needed to clarify these connections. 


\acknowledgments

We are grateful to Dr. Sergiu Cojocaru for useful discussions and for his comments on the manuscript.
This work was supported in part by the Romania-JINR collaboration projects, positions 22, 23, 24, 27, Order 322/21.05.2018 (IFIN-HH), and position 24, Order 323/21.05.2018 (JINR).
DVA was supported by the ANCS project PN18090101/2018.

%

\end{document}